\documentclass[]{emulateapj}

\newcommand{\igr}     {{IGR~$16358-4726$}} 
\newcommand{\fluxu}   {{ergs s$^{-1}$ cm$^{-2}$}} 
\newcommand{\rxte}    {{\it RXTE}}
\newcommand{\chandra} {{\it Chandra}}
\newcommand{\integral} {{\it INTEGRAL}}
\newcommand{\sax}     {{\it BeppoSAX}}
\newcommand{\asca}    {{\it ASCA}}

\newcommand{\sgr}     {SGR~$1627-41$}
\newcommand{\etal}    {{\it et al.~}}

\begin{document}

\title{The Peculiar X-ray Transient \igr}

\author{S.~K.~Patel\altaffilmark{1}, C.~Kouveliotou\altaffilmark{2,3}, 
A.~Tennant\altaffilmark{2}, P.~M.~Woods\altaffilmark{3}, 
A.~King\altaffilmark{4},  P.~Ubertini\altaffilmark{5}, 
C.~Winkler \altaffilmark{6}, T.~J.~-L.~Courvoisier\altaffilmark{7}, 
M.~van der Klis\altaffilmark{8}, S.~Wachter\altaffilmark{9},
B.~M.~Gaensler\altaffilmark{10}, C.~J.~Phillips\altaffilmark{11}}

\email{sandeep.patel@nsstc.nasa.gov}

\altaffiltext{1} {National Research Council Fellow, NSSTC, SD-50, 320
Sparkman Drive, Huntsville, AL 35805, USA}
\altaffiltext{2} {NASA/Marshall Space Flight Center, NSSTC, SD-50, 320
Sparkman Drive, Huntsville, AL 35805, USA}
\altaffiltext{3} {Universities Space Research Association, NSSTC,
SD-50, 320 Sparkman Drive, Huntsville, AL 35805, USA}
\altaffiltext{4} {University of Leicester, Dept. of Physics and Astronomy, Leicester, LE1 7RH United Kingdom}
\altaffiltext{5} {Istituto di Astrofisica Spaziale e Fisica Cosmica, CNR,
Via Fosso del Cavaliere 100, 00133 Rome, Italy}
\altaffiltext{6} {ESA-ESTEC/SCI-SD, Keplerlaan 1, 2201 AZ Noordwijk,
  The Netherlands}
\altaffiltext{7} {ISDC, 16, ch. d'Ecogia, CH-1290 Versoix \& Geneva 
Observatory, ch des Maillettes, 51, CH-1290 Sauverny, Switzerland}
\altaffiltext{8} {Astronomical Institute ``Anton Pannekoek''and Center
for High Energy Astrophysics, University of Amsterdam, Kruislaan 403,
1098 SJ Amsterdam, The Netherlands}
\altaffiltext{9} {SIRTF Science Center, Caltech M/S 220-6, 1200 E. 
California Blvd., Pasadena CA 91125, USA}
\altaffiltext{10} {Harvard-Smithsonian Center for Astrophysics, 60 Garden
Street MS-6, Cambridge, MA 02138}
\altaffiltext{11} {Australia Telescope National Facility, CSIRO PO Box 76, 
Epping, NSW 1710, Australia ; Bolton Fellow}

\begin{abstract}

The new transient \igr\ was discovered on 2003 March 19 with
\integral. We detected the source serendipitously during our 2003
March 24 observation of \sgr\ with the \chandra\ X-ray observatory at
the $1.7 \times 10^{-10}$ \fluxu\ flux level ($2-10$ keV) with a very
high absorption column ($N_{\rm H}=3.3 \times 10^{23}$~cm$^{-2}$) and
a hard power law spectrum of index 0.5(1). We discovered a very strong
flux modulation with a period of 5880(50)~s and peak-to-peak pulse
fraction of 70(6) \% ($2-10$ keV), clearly visible in the
x-ray data. The nature of \igr\ remains unresolved. The only neutron
star systems known with similar spin periods are low luminosity
persistent wind-fed pulsars; if this is a spin period, this transient
is a new kind of object. If this is an orbital period, then the system
could be a compact Low Mass X-ray Binary (LMXB).

\end{abstract}


\slugcomment{To be submitted to ApJ Letters - Version of \today}

\section{Introduction}
Ten new hard and relatively faint X-ray transients have been detected
in the last few months during galactic plane scans by the IBIS/ISGRI 
detector (Ubertini \etal 2003; Lebrun \etal 2003) on the \integral\ Observatory
(Winkler \etal 2003). Their unusual spectral hardness has led to
suggestions that these sources belong to a group of highly absorbed
galactic binary systems, discovered due to the unique IBIS sensitivity
in higher energy bands than earlier monitoring instruments.  Two of
these sources, IGR~J$16318-4848$ and IGR~J$16320-4751$, were
subsequently observed with the {\it XMM}-Newton satellite. The EPIC-PN
detector spectrum of the former source revealed a very strong Fe
K$_{\alpha}$ emission line and significant K$_{\beta}$ and Ni
K$_{\alpha}$ lines that were variable in time (Matt \& Guainazzi 2003),
whereas the latter exhibited no line features (Rodriguez \etal
2003). The spectral continuum of four out of the five sources
(measured with the Rossi X-ray Timing Explorer (\rxte) or {\it XMM}-Newton)
was best fit with power law functions with photon indices, $\gamma
\sim 1.6$. Finally, all sources showed variations ranging in
timescales of $\sim 100$~s to a few 1000~s, but none showed evidence
for coherent modulations. 
				   
\igr\ was detected (Revnivtsev \etal\ 2003a) with \integral\ on 2003 March 19 during a standard ($\sim 2200$~s) Galactic plane scan as a 50 mCrab source ($15-40$
keV). Preliminary analysis of additional \integral\ scans on 2003
March 22 and 26 indicated a decline in the source flux. Analysis of
archival observations with the Advanced Satellite for Cosmology \&
Astrophysics (\asca) and \sax\ in 1999, revealed a source consistent
with the location of \igr. \sax\ observations in 1998 \& 2000,
however, did not detect the source.

We serendipitously observed \igr\ during a \chandra\ scheduled observation of \sgr\ (Kouveliotou \etal 2003a), and then again with Director's
Discretionary Time (DDT) one month later. In addition we triggered
Target of Opportunity observations with \rxte\ and with the Australia
Telescope Compact Array (ATCA); the analysis of the \rxte\ data is
reported in detail by Revnivtsev (2003c).  ATCA observed \igr\ over a
10.5~hr period beginning at UT 11:53 on 2003~April~13.  The total
on-source integration time was 5~hr. No source was detected at the
position of the X-ray source at an observing frequency of 4.8 GHz,
with a $5\sigma$ upper imit of 0.9~mJy. We report in Sections 2, 3 and
4 the results of our analysis of the archival and current X-ray
observations and discuss in Section 5 the implications of these
results for the nature of \igr.

\section{Archival X-ray Observations of \igr}

The field containing \igr\ was observed with \asca\ on 1999 February
26-28 during a scheduled observation of \sgr. The transient was only
in the (wider) field of view of the two \asca/Gas Imaging
Spectrometers (GIS).  We combined the GIS data and grouped the
resulting spectrum  in energy bins that contained at least 25 events
each. Using XSPEC (v11.2; Arnaud 1996) we fit an absorbed single power
law (PL) model, which gave an acceptable fit ($\chi^2/\nu=56.1/58$). The best fit column density and photon power law index are, $N_{\rm H}=20^{+10}_{-7} \times 10^{22}$ cm$^{-2}$ and
$\gamma=1.3^{+0.9}_{-0.8}$, respectively.  We derive an unabsorbed
flux of $2.4\times10^{-12}$ ergs~cm$^{-2}$~s (2$-$10~keV), consistent
with the analysis of Revnivtsev \etal (2003b).

\begin{table}[!ht]
\begin{center}
\label{saxtab}
\caption{\asca\ \& \sax\ Observations of \igr.} 
\begin{tabular}{cccccc} \hline \hline
     & \asca/GIS    & \multicolumn{2}{c}{$2-10$ keV Flux$^{a}$}  \\
Date & Exposure Time (ks) & Absorbed & Unabsorbed \\
\hline
26-28 Feb 1999 &  69  &  $11$  & $24$ \\ \hline \hline
\vspace{3pt}
     & \sax\ MECS/LECS     & \multicolumn{2}{c}{$2-10$ keV Flux}  \\
\hline
06 Aug 1998$^{b}$ & 45/ 21 & $< 6$& $< 13$ \\
16 Sep 1998$^{b}$ & 30/ 12 & $< 9$& $< 21$  \\
08 Aug 1999       & 34/ 80 & $10(2)$ & $ 22(4)$ \\
26 Sep 2000$^{b}$ & 61/ 21 & $< 3$& $< 7$  \\
\hline\hline
\multicolumn{6}{c}{
\parbox{3.5in}{
\vspace{0.2cm}
$^{a}$ Flux is in units of $10^{-13}$ \fluxu, $^{b}$ 90\% upper limit}}
\end{tabular}
\end{center}
\end{table}

\sax\ observed the \sgr\ field four times (Table 1). Using XSELECT
(v2.1) we extracted spectra from circular regions centered on \igr\
with radii of $4\arcmin$ and $6\arcmin$ for the LECS and MECS,
respectively. We use the standard response files provided by the \sax\
Data Archive at HEASARC. The source was not detected during the
observations in 1998 or 2000 and we derived a $90\%$ upper limit for
the unabsorbed flux (Table 1). We detect the source
during the 1999 observations, albeit very faint. We have used the
spectral parameters derived with the \asca\ data to fit the \sax\
source counts and find a flux value that is  slightly higher than the
one reported by Revnivtsev \etal (2003b). In fact, the true flux may
be even higher, when one takes into account the partial source
blockage due to the support structure for the MECS detector
windows.   

\section{\chandra\ Observations of \igr}

We observed the transient for 25.7~ks with the
Advanced CCD Imaging Spectrometer (ACIS) on \chandra\ on 2003 March 25 (Obs1). We used the
CIAO v2.3 and CALDB v2.21 software for all data analysis and
processing tasks. Details of the observation and data reduction
techniques are presented in Kouveliotou \etal (2003b); an accurate
($\lesssim0\arcsec.3$) location of the source is given in Wachter
\etal (2003). During this observation \igr\ fell off axis on the
(front illiuminated CCD) ACIS-S2. The source was very bright, however,
the pile-up effect was negligible due to the significant broadening of
the point spread function (PSF), enabling reliable spectral
measurements (\S 3.2). We observed the source again (Obs2)
on 2003 April 21 for 47 ks; we selected the continuous clocking (CC) mode to
mitigate any pile-up and to exploit the mode's 2.85~ms time
resolution.

\subsection{Timing Analysis}

Strong pulsations with a period $\sim$100 min are directly visible in
the light curve (Figure 1; upper panel) during Obs1 (Kouveliotou \etal
2003a).  We measured the pulse period
by first maximizing the signal-to-noise in the light curve by
selecting photon energies in the range $2-10$ keV and by correcting
the arrival times to the Solar System barycenter. Next, we
constructed a template pulse profile from the full data set using a
trial period determined from an epoch fold search.  The data were
split into 4 roughly equal time segments and folded on this trial
period.  We compared the relative phases of the four segments to the
pulse template and finally we fitted them to a linear trend, which
yielded the correction to our trial period.  After a series of
iterations, we find the best fit period for Obs1 to be 5880(50) s. The
pulse profile has a dominant sinusoidal component with  significant
substructure and varies as a function of energy (Figure 2; left
panel).  A portion of the substructure can be attributed to a
quasi-periodic oscillation feature at $\sim 3\times 10^{-3}$ Hz. The
pulse fraction is constant with energy across the \chandra\ bandpass;
the $2-10$ keV rms value is 36.6(8)\% (70(6)\% peak-to-peak).

During Obs2 the source was detected at a
lower intensity, a factor of $\sim15.5$ fainter in flux ($2-10$
keV).  In addition, a visual inspection of the light curve shows
substantially more flickering than present in Obs1 (Figure 1; lower
panel).  The Fourier power spectrum shows the 100 min pulsations at
much weaker significance superposed upon an excess of red noise.
There are no other coherent high frequency signals out to 175~Hz.
Using the same technique as before, we measure a pulse period of
$5860(30)$~s, consistent with the period derived from Obs1. The folded
pulse profile during Obs2 (Figure 2; right panel) is significantly
different than what it was four weeks earlier; the pulse shape is more
complex with more power in the higher harmonics. Despite the changes
in pulse shape, the pulse fraction has remained high and constant at $39(2)\%$ rms
($2-10$ keV). We performed a cross-correlation of the high and medium energy (Figure 2) pulse profiles of Obs2 and find a soft lag of $\sim110$s; a lag was not detected in a preliminary analysis of Obs1. A refined analysis will be reported elsewhere (Patel \etal in preparation).

\begin{figure}   
\label{fig1}
\figurenum{1}
\epsscale{1.0}
\includegraphics[angle=-90,scale=0.365]{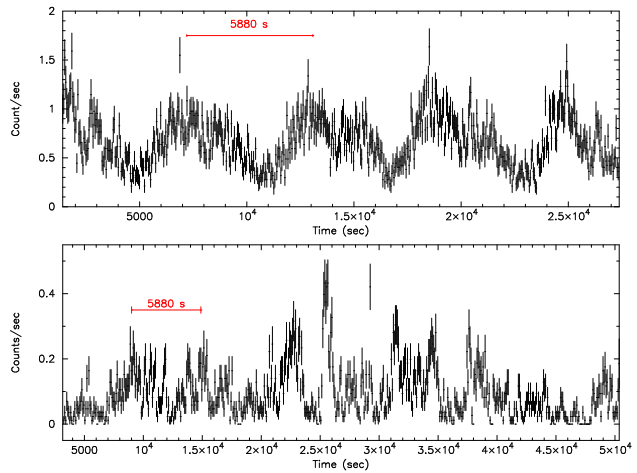}
\caption{{\sl Top panel}:\chandra\ ACIS-S2 lightcurve ($1.0-10.0$~keV) of
\igr\ in an active state on 2003~March~24. The modulation at
5880(50)~s is clearly seen. {\sl Bottom panel}:\chandra\ ACIS-S3
lightcurve of the same source on 2003~April~21. Note the different duration of the
x-axis in the two panels.}
\vspace{-0.5cm}
\end{figure} 

\begin{figure}[h]
\label{fig2}
\figurenum{2}
\begin{center}
\vspace{-0.8cm}
\includegraphics[scale=0.6]{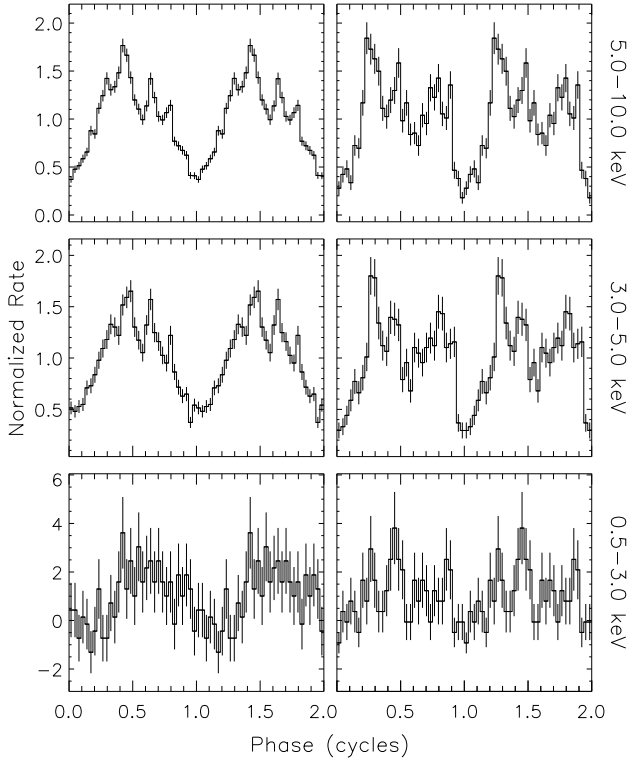}
\vspace{-0.5cm}
\end{center}
\caption{Folded light curves of \igr\ in an active state taken with
  the \chandra/ACIS (in three energy bands) on 2003 March 24 ({\sl
    Left panel}) and 2003 April 21 ({\sl Right panel}).}
\vspace{-0.5cm}
\end{figure} 

\subsection{Spectral analysis}

\igr\ is detected significantly off-axis ($9.7\arcmin$) during Obs1,
necessitating a large extraction radius for the source spectrum to
account for the significant PSF distortion of the source and for the
scattering of the hard photons of the source spectrum. We have, thus,
collected source events from a circular region of $1.5\arcmin$ radius.
We extracted a background spectrum from a source free region of
$3.0\arcmin$ radius and on the same CCD. Finally, we grouped the data
into energy channels that contained at least 25 events each. 

We report here results on our phase-averaged fits to the data. During
Obs1 the spectrum includes a highly significant Fe line feature and is
hard and highly absorbed. We have initially fitted the data with a PL
+ a Gaussian line model; the best fit parameters are given in Table 2, and the spectral fit is shown in Figure 3. The 2003 \chandra\ flux is $\sim70$ times higher than the \asca\ flux of 1999. We have also fitted the data with an absorbed thermal blackbody (BB)
model with a line (Table 2); the fit statistic is slightly
improved. Finally, we have tested two-component (continuum) models of PL + BB and PL + PL; we also find acceptable fits. However, the additional (PL) component in
each case is very soft and steep, mostly contributing to a very narrow spectral range of the continuum below 3 keV, where the signal is very low.  We conclude that the single model
fits (PL or BB) better describe the data. 

During Obs2, the spectrum is also well
described by an absorbed PL. We find that the
absorption column has decreased significantly since Obs1, but is still
in agreement with the best fit column density measured by \asca,
suggesting that at least part of the column is due to local
absorption.  Furthermore, we find that the Fe line feature has faded
beyond detection in Obs2. The upper limit we measure on the line
equivalent width suggests that the flux in the line has faded more
than the continuum flux; the latter is now a factor of 15.5 lower than
Obs1. Figure 3 also shows the PL spectral fit result of Obs2. The
spectral parameters are listed in Table 2, together with the results
of an absorbed BB model fit.

Finally, we have fitted the two \chandra\ observations simultaneously
using an absorbed PL $+$ Fe line model. Here we linked the galactic
absorption and Fe line component and added an additional intrinsic
absorption component; the latter was free to vary for Obs1 but assumed
to be zero for Obs2. With $\gamma$ free, we derive galactic and
intrinsic (Obs1) column densities and photon power law indices
consistent with the individual fit values above (Table 2). In a
similar fashion, we simultaneously fit both observations using an
absorbed BB $+$ Fe line model.  With $N_{{\rm H},gal}$ and line
parameters linked and $kT_{BB}$ free, we find galactic and intrinsic
(Obs1) column densities and temperatures also consistent with the
previous (not linked) values (Table 2). 

\begin{figure}   
\label{fig:chandspec}
\figurenum{3}
\epsscale{1.0}
\includegraphics[angle=-90, scale=0.365]{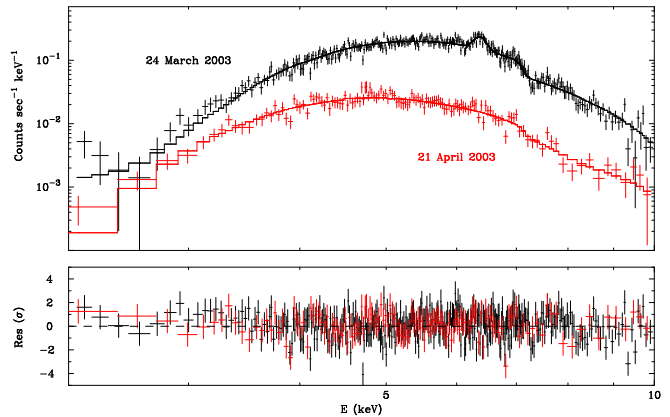}
\caption{\chandra\ ACIS-S3 spectrum of \igr\ in an active state on 24
  March 2003 (top curve) and again on 21 April 2003 (bottom curve). In
  both observations we find that a highly absorbed single power law
  model gives the best fit to the data. A Fe K$_\alpha$ line feature
  clearly seen at 6.4~keV during the 24 March observation had faded
  below the detection capability of \chandra\ during the second
  observation.}
\vspace{-0.5cm}
\end{figure} 

\begin{table*}[!ht]
\begin{center}
\caption{\chandra\ spectral fit parameters$^{a}$}
\label{tab:fit}
\begin{tabular}{cccccccccccc} \hline \hline
Model                          & $\Gamma_1$ & $\Gamma_2$ &  $kT_{BB}$ & $N_{H,Gal}$ & $N_{H,int}$ & $E_{line}$ & $EW$ &  $F_x$ (Abs) &  $F_x$ (Unabs)& $F_{\rm line}$ & $\chi^2$/DOF \\ \hline
\multicolumn{12}{c}{OBS1 ONLY}\\\hline
PL$+$Gaussian                  & $0.5(1)$   & ...        & ...        & $3.3(1)$    & ...         & $6.39(1)$ & $0.14(6)$ & $6.9$ & $17.1$  & $0.30$ & $357.5/316$ \\
BB$+$Gaussian                  & ...        & ...        & $3.1(1)$   & $2.90(8)$   & ...         & $6.39(1)$ & $0.13(3)$ & $6.7$ & $14.2$  & $0.27$ & $346.1/316$ \\
PL$_1+$PL$_2+$Gaussian$^{b}$   & $0.8(1)$   & $10(1)$    & ...        & $4.0(2)$    & ...         & $6.38(1)$ & $0.11(3)$ & $6.8$ & $66.2$  & $0.29$ & $337.6/314$ \\
BB$+$PL$_2+$Gaussian$^{c}$     & ...        & $8(1)$     & $2.8(2)$   & $3.5(2)$    & ...         & $6.38(1)$ & $0.11(3)$ & $6.6$ & $28.7$  & $0.26$ & $330.5/314$ \\\hline
\multicolumn{12}{c}{OBS2 ONLY}\\\hline
PL$+$Gaussian$^{d}$            & $0.8(2)$   & ...        & ...        & $2.0(1)$    & ...         & ... & ...  & $0.6$ & $1.1$  & $< 0.02$ & $124.3/136$ \\
BB$+$Gaussian$^{d}$            & ...        & ...        & $2.6(2)$   & $1.6(1)$    & ...         & ... & ...  & $0.5$ & $0.9$  & $< 0.02$ & $123.5/316$ \\\hline
\multicolumn{12}{c}{OBS1$+$OBS2$^{e}$}\\\hline
PL$+$Gaussian$^{d}$            & $0.5(1), 0.7(2)$ & ...  & ...        & $2.0(1)$    & 1.3(2)      & $6.39(1)$ & $0.14(6)$ & $6.9$,$0.6$ & $17.1$,$1.1$  & $0.24$,$< 0.02$ & $478.8/448$ \\
BB$+$Gaussian$^{d}$            & ...     & ...  & $3.1(2)$, $2.6(2)$  & $1.6(1)$    & 1.3(1)      & $6.39(1)$ & $0.13(3)$ & $6.7$,$0.5$ & $14.2$,$0.9$  & $0.21$,$< 0.02$ & $466.5/448$ \\
\hline\hline
\multicolumn{12}{c}{
\parbox{7.0in}{
\vspace{0.2cm}

{\sc Notes:}  Uncertainties are 68\% confidence limits. All models
include galactic absorption $^{a}$ $kT_{BB}$, $E_{line}$ and EW in
keV, $N_{\rm H}$ in $10^{23}$ cm$^{-2}$, fluxes are in $10^{-11}$
\fluxu  $2-10$ keV) $^{b}$ Double power law $+$ Gaussian line model
fits. PL$_1$ component constitutes $30\%$ of the unabsorbed flux $^{c}
$PL$_2$ corresponds to the additional power law model component in
this model. BB component constitutes $56\%$ of the unabsorbed flux
$^{d}$ Line Flux: $90\%$ flux upper limit for a line at 6.4 keV with
similar intrinsic width as the one in Obs1 $^{e}$ Data from Obs1 and
Obs2 are fit simultaneously. The pairs of values represent the
spectral parameters from Obs1 and Obs2 respectively.}}
\end{tabular}
\end{center}
\vspace{-0.6cm}
\end{table*}

Further, we performed phase-resolved
spectroscopy in five phase-bins with the data of Obs1 using a PL $+$ a Gaussian line
model.  We find that while the column density remains
constant, the spectral index rises slightly during the ascending part
of the pulse profile; the spectrum is marginally softer during the
pulse decay. The 6.4 keV Fe line intensity is constant throughout the
entire pulse. We will report detailed phase-resolved spectroscopy
results for both observations in a future work (Patel \etal in preparation). 

\section{Discussion}

Our \chandra\ observations of the Fe line centroid in the \igr\ spectrum indicate that most likely the source is in a galactic X-ray binary system. If the 1.6 hr period is
orbital, then there are two types of X-ray binary systems which can
have similar periods and X-ray light-curve properties: a Cyg X-3 like
system (4.8~hr period; Milgrom 1976; Parsignault \etal 1977) and
high-inclination accretion disk corona (ADC) systems such as
4U~$1822-37$ (5.57~hr period; White \etal 1981). Like both these
sources \igr\ in outburst is showing a smoothly varying X-ray light-curve;
the depth of the modulation is not indicative of an eclipse and there
is no phase dependent absorption. All three spectra are similar, well
fit with a (hard) power law with a low energy cut off at $\sim4$~keV
and with large, broad iron lines at $\sim6.4$~keV. The source
luminosities are also within the same range of $\sim10^{36}$~erg/s
(assuming a distance for \igr\ of 10~kpc).

It is, however, questionable if a He star with a strong stellar wind
such as in Cyg X-3 could fit a system with an orbital period of
1.6~h. More likely, the system is an LMXB such
as 4U~$1822-37$; the questions that immediately arise are where did
the extreme absorption detected in the spectrum of \igr\ originate if
not from the stellar wind of a massive companion, and is a low mass
companion sufficient to produce the Fe line?  One expects systems with
such periods to be transient (King, Kolb \& Burderi 1996) as the
average mass transfer rate would probably be quite low. The `outburst'
accretion rate could nevertheless be high, thus, if the ADC
interpretation is right, the true luminosity of the object is much
higher than observed. It is just conceivable that this actually causes
the intense absorption, if the accretion rate is highly
super-Eddington for the duration of the outburst. This would produce a
dense outflowing wind, as is probably seen in some of the Ultra
Luminous X-ray sources (Fabbiano \etal 2003), which could give the
absorption. Similar super-Eddington outflows, with extremely high
columns, are seen in quasars inferred to be accreting at
super-Eddington rates (Pounds \etal 2003; King \& Pounds 2003). 
Sources like this or with even higher columns could easily have been 
missed by earlier satellites, so there could be a whole class of 
them.

Finally, we would like to note that if \igr\ is a LMXB, its orbital
period would be in the range of those of the ms pulsar subclass, only
five of which have been found. No ms pulsations were detected so far
and unless the source becomes active again, current instrumentation
does not afford ms pulsation searches. If on the other hand the 1.6 hr
period is a neutron star spin period, the source is similar to
2S~$0114+650$, which has a spin period of $\sim2.7$ hr (Finley,
Belloni \& Cassinelli 1992) and an Fe line at 6.4 keV (Yamauchi \etal
1990); by analogy to other slow accreting pulsars we would expect the
magnetic field of the neutron star in \igr\ to be of order
$10^{12}$~Gauss.

\acknowledgements {We are grateful to H. Tananbaum for the generous
  and prompt allocation of DDT, and to the \chandra\ science support
  center for the rapid data processing and delivery. S. P., C. K., and
  P. W. acknowledge support from NASA grant NAG5-9350, and SAO grant
  GO1-2066X. S. P. also acknowledges support from the NAS/NRC
  Fellowship. The Australia Telescope is funded by the Commonwealth of
  Australia for operation as a National Facility managed by CSIRO.}

\end{document}